\documentstyle[prd,aps,epsf,floats]{revtex}
\begin{document}


\title{Gravitational production of gravitinos}
\author{Martin Lemoine}
\address{DARC, UMR--8629, CNRS,
Observatoire de Paris-Meudon, F-92195 Meudon C\'edex, France\\
Email: Martin.Lemoine@obspm.fr}

\date{\today}
\maketitle

\begin{abstract} We calculate the number density of helicity $\pm3/2$ 
gravitinos produced out of the vacuum by the non-static gravitational
field in a generic inflation scenario. We compare it to the number
density of gravitinos produced in particle interactions during 
reheating.


\end{abstract}



\section{Introduction} It was realized early on~\cite{W82} that the
gravitino could pose a serious cosmological problem in the context of a
hot Big-Bang, if it were once in thermal equilibrium. An unstable
gravitino, for instance, would decay in the post-Big-Bang
nucleosynthesis era, if its mass $m_{3/2}\lesssim10^4\,$GeV, and the
entropy produced would ruin the successes of Big-Bang nucleosynthesis. 
Similarly, if the gravitino were stable, as {\it e.g.}, in
gauge mediated supersymmetry breaking, its energy density would eventually
overclose the Universe, if its mass $m_{3/2}\gtrsim2\,$keV. A
solution to this problem was brought forward in Ref.~\cite{ELN82}: if
inflation took place, the gravitinos present at the time of
Big-Bang nucleosynthesis were created during reheating, in an
abundance possibly much smaller than that corresponding to thermal
equilibrium. Cosmological constraints on their abundance could then be
turned into useful upper limits on the reheating temperature $T_{\rm
R}$~\cite{NOS83}, typically  $T_{\rm R}\lesssim 10^8-10^{10}\,$GeV, for an
unstable gravitino with $m_{3/2}\lesssim
3\times10^3\,$GeV, or $T_{\rm
R}\lesssim10^2-10^{10}\,$GeV, for a stable gravitino with $1\,{\rm
keV}\lesssim m_{3/2}\lesssim1\,{\rm GeV}$.

These studies assume that the gravitino abundance has been
exponentially suppressed during inflation, and that gravitinos were
only created in particle interactions during reheating. However,
particles can be produced out of the vacuum in a non-static
gravitational background if their coupling to the gravitational field
is not conformal~\cite{BD82}. A well-known case is the production of
gravitational waves or scalar density perturbations during inflation.
As we argue below, a massive gravitino is not conformally invariant in
a Friedmann--Robertson--Walker (FRW) background, and our main
objective is thus to quantify the number density of gravitinos that
can be produced gravitationally during inflation.  We will consider a
generic inflation scenario within $N=1$ supergravity~\cite{O90}, and
briefly discuss the more particular case of pre-Big Bang
cosmology~\cite{GV93}.  The cosmological consequences of gravitino
production during inflation were briefly discussed in
Ref.~\cite{LRS98}.  These authors did not actually study the
gravitational production of gravitinos, and rather focused on the spin
0 and spin 1/2 cases, as they were interested in the problem of moduli
and modulini fields. With regards to the gravitino, they assumed that
one particle would be produced per quantum state for modes with
comoving wavenumber $k\lesssim H_{\rm I}$, where $H_{\rm I}$ denotes
the Hubble scale of inflation. This estimate showed that gravitational
production of gravitinos could pose a cosmological problem if the
energy scale $V^{1/4}$ at which inflation takes place saturates its
observational upper bound, {\it i.e.} $V^{1/4}\sim10^{16}\,$GeV, and in
this respect, it justifies further the present work. Our study is more 
specialized than that of Ref.~\cite{LRS98},
as we concentrate exclusively on the gravitino.  However, it is also
more systematic and more detailed, as we derive and solve the
gravitino field equation to provide quantitative estimates of the
number density of gravitinos produced.  We also examine different
cases for the magnitude and dynamics of the gravitino effective mass
term during and after inflation. Finally, we also study the effect of
a finite duration of the transition between inflation and reheating,
using numerical integration of the field equation. This effect is
important, as this timescale defines the ``degree of adiabaticity'' of
the transition, and indeed the number density of gravitinos produced
is found to be inversely proportional to it.

The study of the conformal behavior of the gravitino also bears
interest of its own, apart from any application to cosmology, and to
our knowledge, the quantization of quantum fields in curved space-time
has been examined for spins 0, 1/2, 1 and 2, but not $3/2$~\cite{BD82}
(although the case of a massless gravitino in a perfect fluid
cosmology was studied in Ref.~\cite{GP79}). In the present work, we
focus on the helicity 3/2 modes of the gravitino. The field equations
and the quantization of the helicity 1/2 modes are indeed more
delicate, due to the presence of constraints. These constraints vanish
identically if supersymmetry is unbroken~\cite{DZ76}; in broken
supersymmetry, these constraints do not vanish, but do not induce any
inconsistency~\cite{DZ77}. As shown below, these constraints apply to
the modes of helicity 1/2, not to those of helicity 3/2, and for this
reason, we leave the problem of the helicity 1/2 modes open for a
further study; nonetheless, we present these field equations and their
constraints. The number density of helicity 3/2 gravitinos produced
during inflation that we derive in this paper should thus be
interpreted as a lower limit. Quite probably, however, the number
density of helicity 1/2 gravitinos should be of the same order as that
of helicity 3/2, and the results correct within a factor of order 2.

This paper is organized as follows. In Section~II, we derive the
gravitino field equation for the $\pm3/2$ helicity modes, and in
Section~III, we calculate the number density of gravitinos produced in
a generic inflation scenario. We summarize our conclusions and briefly
discuss the case of pre-Big-Bang string cosmology in Section~IV.  All
throughout this paper, we use natural units $\hbar = c = m_{\rm
Pl}=1$, where $m_{\rm Pl}\equiv(8\pi G)^{-1/2}$ is the reduced Planck mass.
We note $M_{\rm Pl}\equiv\sqrt{8\pi}m_{\rm Pl}$ the Planck mass.
Furthermore, we restrict ourselves to a FRW background, whose metric
is written as: ${\rm d}s^2=g_{\mu\nu}{\rm d}x^\mu{\rm
d}x^\nu=a^2(\eta)(-{\rm d}\eta^2 + {\rm d}x^2 + {\rm d}y^2 + {\rm
d}z^2)$, where $a(\eta)$ is the scale factor, and $\eta$ denotes
conformal time; the Minkowski metric is written $\eta_{ab}$.  We also
use standard conventions on the derivative of the K\"ahler potential
$G(z_i,z_i^\ast)$ with respect to the scalar components $z_i$ of
chiral superfields: $G^i\equiv\partial G/\partial z_i$,
$G_{i^\ast}\equiv\partial G/\partial z^{i\ast}$.  Other notations,
relative to the Dirac matrices, are given in the Appendix.

\section{Field equation}

We consider the gravitino in a background of a classical FRW spacetime,
in the context of $N=1$ supergravity, and adopt the following
lagrangian density:

\begin{equation}
{\cal L}\,=\,\frac{e}{2}R -
\frac{i}{2}\epsilon^{\mu\nu\rho\sigma}\overline\Psi_\mu\gamma_5\gamma_\nu
D_\rho\Psi_\sigma +
\frac{e}{2}e^{G/2}\overline\Psi_\mu\sigma^{\mu\nu}\Psi_\nu
+ {\cal L}_{\rm m}.
\label{eq_lag}
\end{equation}

In this equation, $e$ represents the determinant of the vierbein
$e_\mu^{{}a}$, $R$ denotes the Ricci scalar, $\Psi_\mu$ the gravitino
field, and ${\cal L}_{\rm m}$ represents external matter, more
specifically the scalar fields whose dynamics drive the evolution of
the background metric: we  neglect the matter gauge and fermion fields.
The gravitino covariant derivative $D_\rho$ is defined as:

\begin{equation} 
D_\rho=\partial_\rho + \frac{1}{4}\omega_\rho^{{}ab}\sigma_{ab}
- \frac{1}{4}\gamma_5\lambda_\rho, 
\label{eq_covd} 
\end{equation}
where $\omega_\rho^{{}ab}(e)$ is the torsion tensor, in which we do not
include $\Psi$ torsion, since we neglect the backreaction of the
gravitino on the metric. We included in this covariant derivative the
K\"ahler connection $\lambda_\rho=K^i\partial_\rho z_i -
K_{i^\ast}\partial_\rho z^\ast_i$, where $K$ denotes the K\"ahler
function. The gravitino is also coupled to matter through the K\"ahler
potential $G(z,z^\ast)=K(z,z^\ast) + \ln\left(|W(z)|^2\right)$, with
$W$ the superpotential. This term gives rise to an effective mass for
the gravitino, which we write as $m$: $m\equiv e^{G/2}$.  The gravitino
field equation can be written in the compact notation~\cite{DZ77}:

\begin{equation}
R^\mu=\epsilon^{\mu\nu\rho\sigma}\gamma_5\gamma_\nu{\cal D}_\rho\Psi_\sigma=0,
\label{eq_rmu}
\end{equation}
where  ${\cal D}_\rho\equiv D_\rho + \frac{1}{2}m\gamma_\rho$.
As is well-known~\cite{DZ77,DZ76}, a consistency condition can be
obtained by taking the divergence of Eq.~(\ref{eq_rmu}), ${\cal D}_\mu
R^{\mu}$=0, which leads, after some manipulations, to:

\begin{equation}
\left[
3m^2\gamma^\nu - G^\nu_{\,\mu}\gamma^\mu + 2\partial_\mu
m\sigma^{\mu\nu} + 2m\gamma_5\lambda_\mu\sigma^{\mu\nu}
\right]\Psi_\nu \,=\,0,
\label{eq_cons}
\end{equation}
where $G_{\mu\nu}$ is the Einstein tensor, symmetric in the absence of
torsion. In this equation, we did not include a term of the form
$\epsilon^{\mu\nu\ldots}\partial_\mu\lambda_\nu\ldots$, since it
vanishes in a homogeneous and isotropic background.  

We now define: ${\cal R}_\mu = R_\mu - \frac{1}{2}\gamma_\mu \gamma^\nu
R_\nu$, and rewrite the field equation Eq.(\ref{eq_rmu}), in an
equivalent way, as ${\cal R}_\mu=0$:

\begin{mathletters}
\label{eq_R}
\begin{eqnarray}
{\cal R}_0 & = &
\left(\gamma^\nu\partial_\nu + m + \frac{3{\cal H}}{2}\gamma^0
+\frac{1}{2}\gamma_5\gamma^0\lambda_0\right)\Psi_0 -
\left(\partial_0 - \frac{1}{2}m\gamma_0 + {\cal H}
+\frac{1}{2}\gamma_5\lambda_0\right)\gamma^\nu\Psi_\nu
=0, 
\label{eq_r0}\\
{\cal R}_i & = & \left(\gamma^\nu\partial_\nu + m +
\frac{{\cal H}}{2}\gamma^0 + 
\frac{1}{2}\gamma_5\gamma^0\lambda_0\right)\Psi_i -
\left(\partial_i - \frac{1}{2}m\gamma_i + 
\frac{{\cal H}}{2}\gamma_i\gamma^0 \right)
\gamma^\nu\Psi_\nu + 
{\cal H}\gamma_i\Psi^0 =0,
\label{eq_ri}
\end{eqnarray}
\end{mathletters}
where ${\cal H}\equiv a'/a$. Note that in a FRW background, $\lambda_i=0$,
and $\partial_j\lambda_\mu=0$. Another integrability condition can
be obtained from the difference $\gamma^0{\cal R}_0 - \gamma^i{\cal R}_i$:

\begin{equation}
g^{ij}\partial_i\Psi_j=(\gamma^i\partial_i - m + {\cal H}\gamma^0)\gamma^j
\Psi_j.
\label{eq_cons2}
\end{equation}
Equation~(\ref{eq_R}) and the constraints Eqs.~(\ref{eq_cons}) and 
(\ref{eq_cons2}) form the system of field equations for the gravitino. 

We now perform a standard decomposition of the gravitino field
operator.  We rescale the gravitino field, and write:
$\Psi_\mu(x)=a(\eta)^{-3/2}e_\mu^{{}c}\hat\Psi_c(\eta,\bbox{k})e^{i\bbox{kx}}$;
we recall that we reserve latin indices $a,b,c\ldots$, or a hat, if
confusion could arise, for Lorentz indices, and
$e_\mu^c=a(\eta)\delta_\mu^c$. Note also that $\Psi_\mu(x)$ transforms
with a conformal weight $1/2$, and $\hat \Psi_a(x)$ transforms with a
conformal weight $3/2$. Then, we decompose the spatial part of
$\hat\Psi_c(\eta,\bbox{k})$ into helicity eigenstates~\cite{M95}:

\begin{equation}
\hat\Psi_c(\eta,\bbox{k})=\mathop{\sum_{m=L,+,-}}_{s=\pm}
C_{1,1/2}(m+\frac{s}{2};m,\frac{s}{2})
\epsilon^m_c(\bbox{k})\psi_{ms}(\eta,\bbox{k}),
\label{eq_dec}
\end{equation}
where $C_{1,1/2}(m+\frac{s}{2};m,\frac{s}{2})$ is a Clebsch-Gordan
coefficient, and $\bbox{\epsilon}^{\rm L}$, $\bbox{\epsilon}^+$, and
$\bbox{\epsilon}^-$ are polarization vectors.  They satisfy:
$\bbox{\epsilon}^{s\ast}\bbox{\epsilon}^{s'}=\delta_{ss'}$, with
$s={\rm L}, +, -$; in particular, $\bbox{\epsilon}^{\rm L}$ is
parallel to $\bbox{k}$, $\bbox{\epsilon}^+$ and $\bbox{\epsilon}^-$
are tranverse to $\bbox{k}$, and
$\bbox{\epsilon}^{+\ast}=\bbox{\epsilon}^-$.  Similarly, the spinors
$\psi_{ms}$ are helicity eigenstates of the helicity operator ${\rm
diag}\left(\bbox{\epsilon}^{\rm L}\bbox{\sigma}, \bbox{\epsilon}^{\rm
L}\bbox{\sigma}\right)$. More specifically, each spinor
$\psi_{ms}(\eta,\bbox{k})$ is written in terms of a Weyl spinor
$\chi_s(\bbox{k})$ of helicity $s/2$, {\it i.e.} such that
$\bbox{\epsilon}^{\rm L}\bbox{\sigma}\,\chi_s\,=\,s\,\chi_s$, and mode
functions $h_{ms}(\eta,k)$ and $g_{ms}(\eta,k)$, where
$k\equiv|\bbox{k}|$, following the notations of the Appendix.  In this
decomposition, the vector-spinors $\bbox{\epsilon}^{\rm L}\psi_{{\rm
L}+}$, $\bbox{\epsilon}^{\rm L}\psi_{{\rm L}-}$,
$\bbox{\epsilon}^+\psi_{+-}$, $\bbox{\epsilon}^-\psi_{-+}$, and
$\Psi_0$ form the helicity $\pm1/2$ components of $\hat\Psi_c$, while
$\bbox{\epsilon}^+\psi_{++}$ and $\bbox{\epsilon}^-\psi_{--}$ are the
helicity $\pm3/2$ components.

The field equation for the helicity $\pm3/2$ modes of the gravitino
can now be extracted from Eq.~(\ref{eq_R}).  To start with, one notes
that the helicity $\pm3/2$ components do not appear in the product
$\bbox{\gamma\Psi}$, because $\bbox{\epsilon}^\pm\bbox{\gamma}$
project out the modes with spinor helicity $\pm$:
$\bbox{\epsilon}^{\pm}\bbox{\sigma}\,\chi_{\pm}=0$, and
$\bbox{\epsilon}^{\pm}\bbox{\sigma}\,\chi_{\mp}=\sqrt{2}\chi_{\pm}$.
Therefore, Eqs.~(\ref{eq_cons}), (\ref{eq_r0}) and (\ref{eq_cons2})
only concern the helicity $1/2$ components, not the helicity $3/2$.  The
field equation for the $\pm3/2$ helicity modes is then obtained by
contracting Eq.~(\ref{eq_ri}) with
$\bbox{\epsilon}^{\mp}\bbox{\hat\gamma}\epsilon^{\mp}_b\eta^{bc}$. This
contraction projects out all terms of helicity $\pm1/2$, because these
are either parallel to $\bbox{k}$, or of the form $\gamma_c\ldots$,
and
$\bbox{\epsilon}^{\mp}\bbox{\hat\gamma}.\bbox{\epsilon}^{\mp}\bbox{\hat\gamma}=
\bbox{\epsilon}^{\mp}\bbox{\epsilon}^{\mp}=0$. Thus the helicity $\pm3/2$
components do not mix with the $\pm1/2$ helicity modes, and their
field equation reads:

\begin{equation}
\left(\hat\gamma^0\partial_0 + i\bbox{\hat\gamma k} + am
+\frac{1}{2}\gamma_5\hat\gamma^0\lambda_0\right)\psi_{ss} = 0,
\,\,\,s=\pm. 
\label{eq_mode3_2}
\end{equation}

Finally, this equation can be rewritten in the usual way as two
systems of two linear and coupled differential equations in the mode
functions $h_{++}(\eta)$, $g_{++}(\eta)$, and $h_{--}(\eta)$,
$g_{--}(\eta)$.  For $m=0$, and zero K\"ahler connection, it is easy
to see that Eq.~(\ref{eq_mode3_2}) is identical to the field equation
for a massless gravitino in Minkowski spacetime, or, in other words, a
massless and uncoupled helicity $3/2$ gravitino is conformally
invariant. 

The gravitino field can now be quantized, following the methods
developed for spin $1/2$ fermions in curved
spacetime~\cite{P79,LRS98}, or, what is similar, for electrons
in an external electromagnetic field~\cite{KESCM92}. Introducing the
shorthand notation:
$\hat\Psi_{c\,\pm3/2}\equiv\epsilon^{\pm}_c\psi_{\pm\pm}$, it can be
checked that the inner product
$\hat\Psi^\dagger_{a\,s}\eta^{ab}\hat\Psi_{b\,s}$, where $s=\pm3/2$,
is conserved by virtue of the field equations. The solutions of
Eq.~(\ref{eq_mode3_2}) are normalized according to:
$\psi^\dagger_{ss}\psi_{s's'}=\delta_{ss'}$, $s,s'=\pm$, and one
obtains at all times:

\begin{mathletters}
\label{eq_norm}
\begin{equation}
\hat\Psi^\dagger_{a\,s}\eta^{ab}\hat\Psi_{b\,s'}=\delta_{ss'},
\label{eq_norm1}
\end{equation}
\begin{equation} 
\hat\Psi^\dagger_{a\,s}(\eta,\bbox{k})\eta^{ab}
\hat\Psi^{\rm C}_{b\,s'}(\eta,-\bbox{k}) = 0,\,\,\,s,s'=\pm3/2,
\label{eq_norm2}
\end{equation} 
\end{mathletters}
where the superscript ${\rm C}$ denotes charge conjugation.  
The helicity$-3/2$ gravitino field operator is written as:

\begin{equation}
\hat\Psi^{(3/2)}_{a}(x) = \int\frac{{\rm d}\bbox{k}}{(2\pi)^{3/2}}
\sum_{s=\pm3/2}
\left[
b(\bbox{k})\hat\Psi_{a\,s}(\eta,\bbox{k})e^{i\bbox{kx}} +
b^\dagger(\bbox{k})\hat\Psi_{a\,s}^{\rm C}(\eta,\bbox{k})e^{-i\bbox{kx}}
\right],
\label{eq_canon}
\end{equation}
where the $b$, $b^\dagger$ are annihilation and creation operators
respectively.  They are related by hermitian conjugation
as the gravitino is a Majorana fermion. Finally, one can relate
field operators  $\hat\Psi^{\rm in}_{a\,s}(\eta,\bbox{k})$ and
$\hat\Psi^{\rm out}_{a\,s}(\eta,\bbox{k})$, that are solutions of the field
equation, and whose boundary conditions are respectively defined
at conformal times $\eta_{\rm in}$ and $\eta_{\rm out}$, by means of a
Bogoliubov transform~\cite{BD82}:

\begin{equation}
\hat\Psi^{\rm out}_{a\,s}(\eta,\bbox{k})  = 
\alpha_{ks}\hat\Psi^{\rm in}_{a\,s}(\eta,\bbox{k})  +  
\beta_{ks} \hat\Psi^{{\rm in}\,\rm C}_{a\,s}(\eta,-\bbox{k}).
\label{eq_Btransform}
\end{equation}

The Bogoliubov coefficients $\alpha_{ks}$ and $\beta_{ks}$ satisfy at all 
times: $|\alpha_{ks}|^2 + |\beta_{ks}|^2=1$, as required for a half-integer
spin field. The occupation number operator for the in quantum state
with momentum $k$ and helicity $s$, $s=\pm3/2$, in the out vacuum, is
then $|\beta_{ks}|^2$, and:

\begin{equation}
|\beta_{ks}(\eta)|^2= \left| h^{\rm in}_{ss}(\eta)g^{\rm
out}_{ss}(\eta) - g^{\rm in}_{ss}(\eta)h^{\rm out}_{ss}(\eta)
\right|^2.
\label{eq_Bbeta}
\end{equation}

 In the following, we solve the field equations for the helicity
$\pm3/2$, and use Eq.~(\ref{eq_Bbeta}) to calculate the number density
of gravitinos produced. 

\section{Gravitational production of spin$-3/2$}

We now assume that the background undergoes an era of inflation,
followed by radiation or matter domination. The magnitude and the
evolution of the gravitino mass term in both epochs are
model-dependent, since the scalar potential $V(z,z^\ast)$ is tied to
the K\"ahler potential in a non-trivial way:

\begin{equation} 
V(z,z^\ast)=
e^{G}\left[G_{i^\ast}(G^{-1})^i_{j^\ast}G^j - 3\right] +
D-{\rm terms}
\label{eq_scalarpot}
\end{equation} 

Nevertheless, it is well-known that scalar fields generically receive
a contribution to their mass of order of the Hubble
constant~\cite{DRT95}, and we adopt this as an ansatz for the
gravitino mass term, {\it i.e.}  $m=\mu_1 H$ during inflation, and
$m=\mu_2 H$ during radiation/matter domination, where $\mu_1$ and
$\mu_2$ are constant parameters, $H$ is the Hubble constant. During
inflation, $H=H_{\rm I}$ is also assumed constant. Note that this
ansatz may be realised rather generically in inflationary
scenarios. For instance~\cite{FKL99}, the superpotential
$\sqrt{\lambda}\phi^3$ gives a potential $\lambda\phi^4$ (albeit in a
global supersymmetry approximation), a gravitino mass
$\sim\sqrt{\lambda}\phi^3/m_{\rm Pl}^2$ (also neglecting K\"ahler
terms), and a Hubble constant $\sim\sqrt{\lambda}\phi^2/m_{\rm
Pl}$. In this model of chaotic inflation, $\phi\sim M_{\rm Pl}$
towards the end of slow-roll, and therefore $m\sim H$. Similarly, for
new inflation types of models, the superpotential $M^2(m_{\rm
Pl}-\phi)^2/m_{\rm Pl}$ gives a scalar potential $\sim M^4$ when
$\phi\ll m_{\rm Pl}$, a gravitino mass $\sim M^2/m_{\rm Pl}$, and a
Hubble constant $\sim M^2/m_{\rm Pl}$.

The quantities $\mu_1$ and $\mu_2$ above can take any value, and,
presumably, $\mu_1\lesssim1$ and $\mu_2\lesssim1$~\cite{O90,S97}. Note
that, strictly speaking, this ansatz is justified as long as
$\mu_{1,2}H>m_{3/2}$, where $m_{3/2}$ denotes the mass of the
gravitino in the true vacuum of broken supersymmetry; provided
inflation takes place at an energy scale $V^{1/4}>10^{11}\,{\rm
GeV}\,(m_{3/2}/10^3\,{\rm GeV})^{1/2}$, this relation should be
satisfied for reasonable values of $\mu_1$ and $\mu_2$. If, however,
$V^{1/4}\ll10^{11}\,{\rm GeV}\,(m_{3/2}/10^3\,{\rm GeV})^{1/2}$, then
according to the adiabatic theorem~\cite{P69,BD82}, the production of
gravitinos will be exponentially suppressed. Nevertheless, for the
sake of completeness, we also present results for this case where $m$
is constant during both inflation and radiation/matter domination.

 For reasons that are similar to the above, one cannot write a generic
K\"ahler connection $\lambda_\rho$ for a generic inflation
scenario. It has actually been argued that if inflation is to proceed
{\it via} the $F-$terms, the K\"ahler function $K$ should not have a
minimal form~\cite{O90,S97}.  Out of simplicity, we thus assume that
this term is zero. This is realized, for instance, in scenarios in
which the dynamical scalar field is real.  Moreover, as we argue in
Section~IV in the case of string cosmology, a non-zero K\"ahler
connection in a homogeneous and isotropic background does not induce
particle creation by itself.  With these assumptions, the differential
equations satisfied by the mode functions read:

\begin{mathletters}
\label{eq_diff1}
\begin{eqnarray}
g_{ss}'' & - & {\cal H}g_{ss}' + \left( k^2 + a^2m^2 - isk{\cal H}\right)
g_{ss} = 0
\label{eq_diff1a}\\
h_{ss} & = & \frac{is}{am}\left(g_{ss}' + iskg_{ss}\right),\,\,\,s=\pm,
\label{eq_diff1b}
\end{eqnarray}
\end{mathletters}
where ${\cal H}$ has been redefined as ${\cal H}\equiv (am)'/(am)$. 
These equations can be solved
in terms of Whittaker functions $W_{\lambda,i\mu_j\alpha_j}(z_j)$ and
$W_{-\lambda,i\mu_j\alpha_j}(-z_j)$, $j=1,2$, where
$z_j\equiv2ik\alpha_j|\eta_{\rm I}|(1 + (1 + \eta/|\eta_{\rm I}|)/\alpha_j)$,
$\lambda=\pm1/2$, and $\eta_{\rm I}$ denotes the conformal time of exit
of inflation: $\eta_{\rm I}=-H_{\rm I}^{-1}$, as we set 
$a(\eta=\eta_{\rm I})\equiv1$. 
The subscript $j=1,2$ correspond to the two eras, $j=1$ for
inflation, and $j=2$ for radiation/matter domination;
 $\alpha_j$ is defined by:  $a(\eta)=(1 + (1 +
\eta/|\eta_{\rm I}|)/\alpha_j)^{\alpha_j}$, {\it i.e.} 
$\alpha_1=-1$ corresponding to de Sitter, and 
$\alpha_2=1,2$ corresponding respectively to 
radiation or matter domination. The in solution is defined as that which
reduces to positive energy plane waves as $\eta\to-\infty$, and the
out solution as that which reduces to positive energy plane waves as
$\eta\to+\infty$. Using the large argument limit of Whittaker functions, one 
obtains~\cite{GR63}:

\begin{mathletters}
\label{eq_Whittaker}
\begin{eqnarray}
g^{(j)}_{++}(\eta) & = & \frac{1}{\sqrt{z_j}}W_{1/2,i\mu_j\alpha_j}(z_j),
\label{eq_Whittakera}\\
h^{(j)}_{++}(\eta) & = & \frac{i\mu_j\alpha_j}{\sqrt{z_j}}
W_{-1/2,i\mu_j\alpha_j}(z_j), \,\,\,j=1,2.
\label{eq_Whittakerb} 
\end{eqnarray} 
\end{mathletters}

In this equation $j=1$ corresponds to the in solution for $\eta<\eta_{\rm I}$,
and $j=2$ corresponds to the out solution. In the radiation or matter 
domination region, the ${\rm in}$ solution reads
$g_{++}=c_1W_{-1/2,i\mu_2\alpha_2}(-z_2) + c_2W_{1/2,i\mu_2\alpha_2}(z_2)$, 
and the coefficients $c_1$ and $c_2$ can be 
obtained by matching $g_{++}$ and $h_{++}$ with 
$g^{(1)}_{++}(\eta<\eta_{\rm I})$ and 
$h^{(1)}_{++}(\eta<\eta_{\rm I})$ in
Eq.~(\ref{eq_Whittaker}) continuously at $\eta=\eta_{\rm I}$.
Finally, the solutions of
helicity $-3/2$ are expressed in terms of the solutions of helicity
$+3/2$: $h_{--}=g_{++}$ and 
$g_{--}=h_{++}$.  
Using Eq.~(\ref{eq_Bbeta}), the asymptotic number of particles produced 
per quantum state $|\beta_{ks}(\eta\to+\infty)|^2$ then reads:

\begin{equation}
|\beta_{ks}(+\infty)|^2= \frac{1}{|z_1||z_2|}\left|
\mu_2\alpha_2 W_{+1/2,i\mu_1}(z_1)W_{-1/2,i\mu_2\alpha_2}(z_2)
+ \mu_1 W_{-1/2,i\mu_1}(z_1)W_{+1/2,i\mu_2\alpha_2}(z_2) \right|^2,
\label{eq_betasol}
\end{equation}
where $z_1=-2ik|\eta_{\rm I}|$, and $z_2=2ik\alpha_2|\eta_{\rm I}|$, and
$\alpha_2=1,2$ depending on whether inflation is followed by radiation or
matter domination. The limits $\mu_1\to0$, or $\mu_2\to0$ are
non-singular, and reduce to the solutions one would obtain in either of
these limits, even though Eq.~(\ref{eq_diff1}) reads differently in these
limits (it decouples into two first-order uncoupled differential equations).

\begin{figure}[ht]
\centering\leavevmode
\epsfxsize=4in
\epsfbox{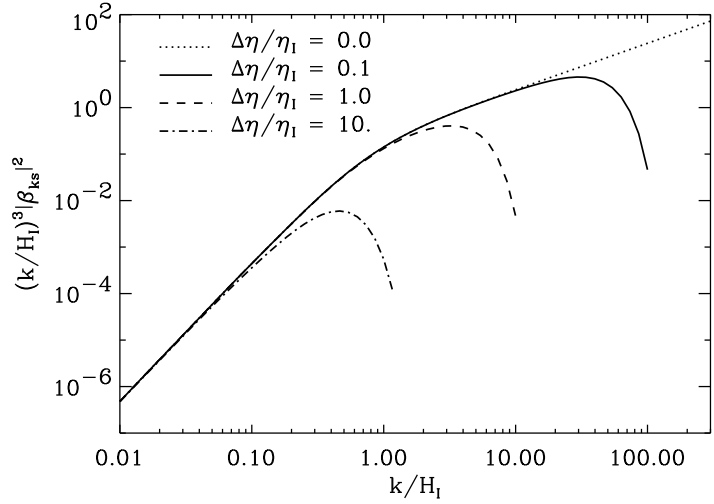}
\bigskip
\caption[...]{Plot of the power spectrum $(k/H_{\rm I})^3|\beta_{ks}|^2$
of the number density of gravitinos per logarithmic wavenumber interval, 
versus $k/H_{\rm I}$, for different transition durations $\Delta\eta$
($\eta_{\rm I}=-H_{\rm I}^{-1}$), as indicated, assuming that
$\mu_1=0$ and $\mu_2=1$, and that matter domination follows inflation. 
The dashed line, which corresponds to $\Delta\eta=0$, is obtained from
the analytical solution in Eq.~(\ref{eq_betasol}); the other curves are
obtained from a numerical integration of the field equation, with a smooth
transition for $a(\eta)$ and $m(\eta)$ between inflation and matter 
domination.}
\label{F1}
\end{figure}

\begin{figure}[ht]
\centering\leavevmode
\epsfxsize=4in
\epsfbox{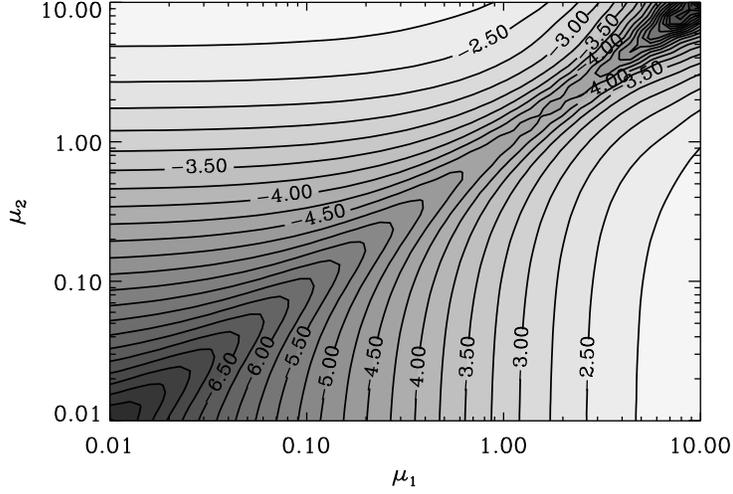}
\bigskip
\caption[...]{Logarithmic (base 10) contour plot of 
$\hat Y_{3/2}\equiv Y_{3/2}(g_{\ast s}/200)(T_{\rm R}/H_{\rm I})^3
(a_{\rm R}/a_{\rm I})^3$,
for $\Delta\eta/|\eta_{\rm I}|=1$, and for a transition into matter 
domination, in the plane $\mu_1-\mu_2$.}
\label{F2}
\end{figure}

\begin{figure}[ht]
\centering\leavevmode
\epsfxsize=4in
\epsfbox{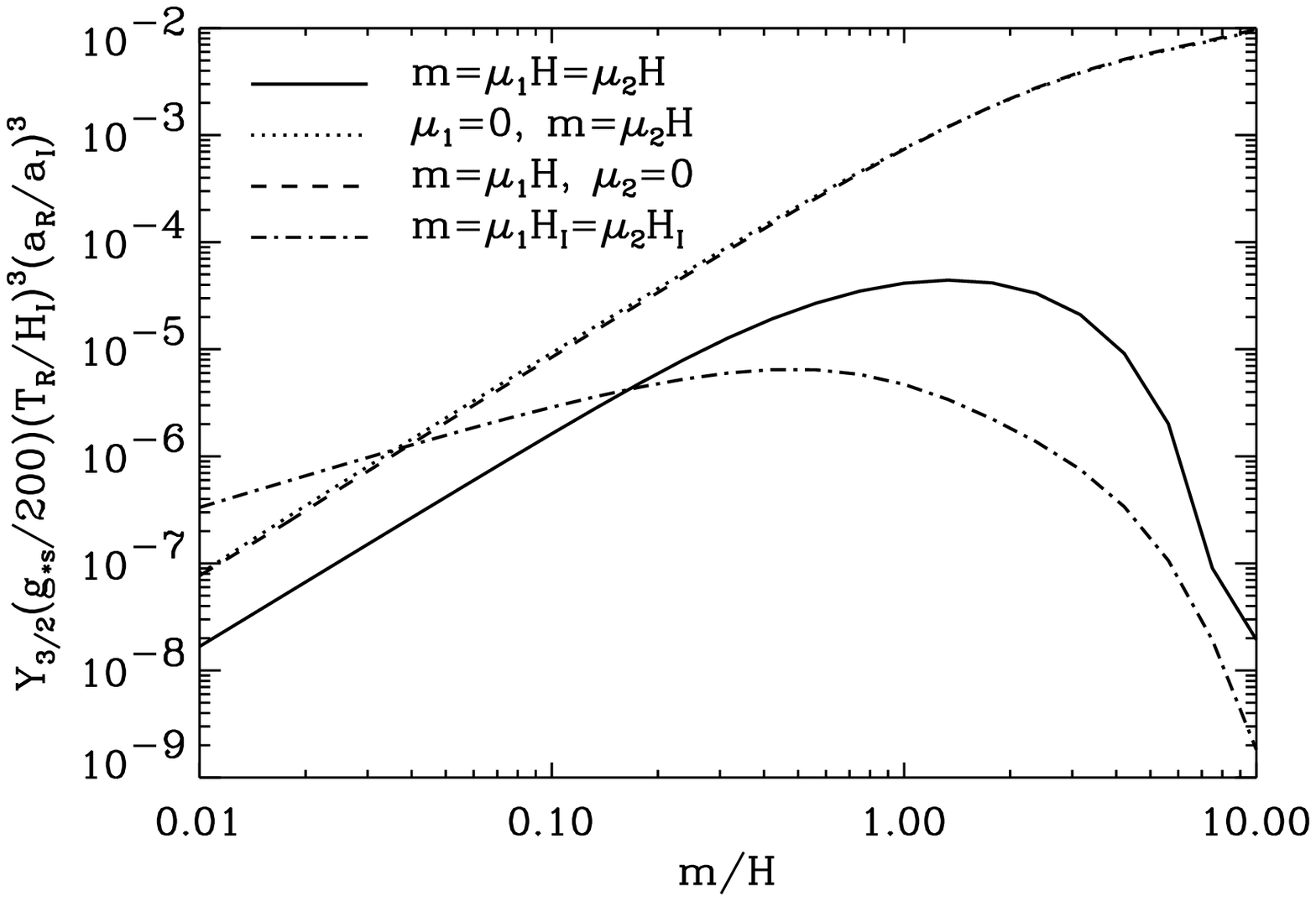}
\bigskip
\caption[...]{Dependence of
$\hat Y_{3/2}\equiv Y_{3/2}(g_{\ast s}/200)(T_{\rm R}/H_{\rm I})^3
(a_{\rm R}/a_{\rm I})^3$ on
the effective mass of the gravitino in different cases: solid line 
$m/H\equiv\mu_1=\mu_2$, dotted line $m/H\equiv\mu_2$ and $\mu_1=0$,
dashed line $m/H\equiv\mu_1$ and $\mu_2=0$, and in dash-dotted line,
as a function of a constant mass term, equal in both inflation and matter 
domination eras. In each case, the transition is operated on a timescale
$\Delta\eta/|\eta_{\rm I}|=1$ into matter domination.}
\label{F3}
\end{figure}

 The quantity of direct interest to us is $Y_{3/2}$, defined as the
ratio of the number density of helicity$-3/2$ gravitinos to the entropy
density at the time of reheating $\eta_{\rm R}$:

\begin{equation}
Y_{3/2}(\eta_{\rm R})=\frac{45}{4\pi^4g_{\ast s}}
\left(\frac{a_{\rm I}H_{\rm I}}{a_{\rm R}T_{\rm R}}\right)^3 
\sum_{s=\pm}
\int{\rm d}{\tilde k} {\tilde k}^2|\beta_{ks}|^2,
\label{eq_density} 
\end{equation}
where $g_{\ast s}$ is the effective number of degrees of freedom,
$g_{\ast s}\simeq229$ for the minimal supersymmetric standard model,
$T_{\rm R}$ is the reheating temperature, 
$a_{\rm R}\equiv a(\eta_{\rm R})$, $a_{\rm I}\equiv a(\eta_{\rm I})$,
and ${\tilde k}\equiv k/H_{\rm I}=k|\eta_{\rm I}|$.

Since $|\beta_{ks}|^2\leq1$ is imposed by Pauli blocking, the integral
in Eq.~(\ref{eq_density}) is dominated by the high wavenumber
modes. As a matter of fact, this integral diverges linearly, since
$|\beta_{ks}|^2\propto k^{-2}$ for $k|\eta_{\rm I}|\gg1$, according to
Eq.~(\ref{eq_betasol}). This divergence is unphysical, and results
from the sudden transition approximation, as is well-known, for
instance, in the case of gravitational waves production during
inflation~\cite{F87,A87}.  The adiabatic theorem~\cite{P69,BD82} indeed
implies that $|\beta_{ks}|^2$ falls off exponentially with $k$ beyond
some cut-off $k_{\rm c}$ (see also Ref.~\cite{CKR99} for a recent
study).  In effect, a numerical integration of the field equation
shows that if the transition between inflation and radiation/matter
domination is sufficiently smooth, an exponential cut-off appears, as
shown in Fig.~(\ref{F1}). Since the integral in Eq.~(\ref{eq_density})
is proportional to $k_{\rm c}$ (provided $k_{\rm c}|\eta_{\rm I}|>1$),
we will use numerical integration of the field equation for
quantitative estimates, and the analytical solution to understand the
behavior in various regimes of $k$, $\mu_1$, and $\mu_2$.

  In Fig.~(\ref{F1}), we compare the analytical (dashed line) and numerical
solutions to $|\beta_{ks}|^2$ in the case where $\mu_1=0$ and $\mu_2=1$, 
and for a transition into matter domination. 
The exact value of the cut-off wavenumber $k_c$ depends
on the duration $\Delta\eta$ of the transition between inflation and matter 
domination~\cite{P69,A87}, as indeed, the ``degree of non-adiabaticity'' of the
transition is inversely proportional to $\Delta\eta$. 
Figure~(\ref{F1}) shows that the analytical solution is an excellent 
approximation to the numerical solution for $k\lesssim k_{\rm c}$, even for
$k_{\rm c}|\eta_{\rm I}|>1$. 
Furthermore, as expected, $k_{\rm c}\propto 1/\Delta\eta$, 
and therefore the number density
in Eq.~(\ref{eq_density}) also scales approximately as $1/\Delta\eta$. Indeed,
as shown in Fig.~(\ref{F1}), the analytical solution for the sudden transition
corresponds to the numerical solution with the cut-off $k_{\rm c}$ cast to
infinity.

  In Fig.~(\ref{F2}), we show a (base 10) logarithmic contour plot of
$Y_{3/2}(g_{\ast s}/200)(T_{\rm R}/H_{\rm I})^3(a_{\rm R}/a_{\rm I})^3$
in the plane $\mu_1,\mu_2$, assuming a transition with 
$\Delta\eta/|\eta_{\rm I}|=1$ into matter domination. A transition into
radiation domination gives similar results. As $\mu_1=\mu_2\to+\infty$, 
the production is exponentially suppressed [see also Fig.~(\ref{F3})],
in agreement with the adiabatic theorem. This can be seen in 
Eq.~(\ref{eq_betasol}), at least in the limit where $k|\eta_{\rm I}|\ll1$, for
which $|\beta_{ks}|^2\propto \exp(-2\pi\mu_1)$. In the limit
$\mu_1\to0$, with $\mu_2$ fixed, one has $Y_{3/2}\propto\mu_1^2$, and similarly
for $\mu_2\to0$, with $\mu_1$ fixed. Notably, $Y_{3/2}\to0$ as $\mu_1\to0$ and
$\mu_2\to0$, since a massless gravitino is conformally invariant.
 Finally, for a fixed $\mu_1$, the
number density  is not exponentially suppressed as $\mu_2\to+\infty$; indeed,
Eq.~(\ref{eq_betasol}) gives $|\beta_{ks}|^2\to1/2$ when
$\mu_1\to0$ and $\mu_2\to+\infty$, for $k|\eta_{\rm I}|<1$. Since 
$k_{\rm c}$ is roughly proportional to max$(\mu_1,\mu_2)$, the number density 
increases linearly with $\mu_2$ in this limit. Note that this does not 
contradict the adiabatic theorem: as $\mu_1\to0$ and 
$\mu_2\to+\infty$, for a fixed $\Delta\eta$, the transition becomes 
increasingly non-adiabatic, and particle production is not suppressed.

  Finally, in Fig.~(\ref{F3}), we show cuts of the previous contour
plot, for $\mu_1=\mu_2$ [diagonal of Fig.~(\ref{F2})], for $\mu_1=0$
[$y-$axis of Fig.~(\ref{F2})], and for $\mu_2=0$ [$x-$axis of
Fig.~(\ref{F2})], in each case for a transition into matter domination
with $\Delta\eta/|\eta_{\rm I}|=1$. For the sake of completeness, we
also include the result of a numerical integration for the case where
the mass of the gravitino is constant and has the same value in both
inflationary and matter dominated epochs, in dash-dotted line. Just
like the case $\mu_1=\mu_2$, it shows exponential suppression as $m\gg
H_{\rm I}$.

The number density of gravitationally produced gravitinos present at
the time of reheating thus depends on several parameters, notably the
effective mass terms during and after inflation, the duration of the
transition from inflation to reheating, and the number of e-folds of
reheating. The yield of gravitinos scales as the inverse of the
transition timescale, which defines the degree of
''non-adiabaticity'', and decreases as $\exp(-3N_{\rm R})$, where
$N_{\rm R}\equiv\ln(a_{\rm R}/a_{\rm I})$ is the number of e-folds of
reheating, during which the gravitinos are diluted. These dependences
make a direct comparison with the number density of gravitinos
produced in particle interactions in reheating slightly delicate.

Let us first isolate the dependence on the mass terms and transition
timescale in the fiducial quantity $\hat Y_{3/2}$, which is defined
through: $Y_{3/2}=\hat Y_{3/2} (g_{\ast s}/200)^{-1}(H_{\rm I}/T_{\rm
R})^3\exp(-3N_{\rm R})$; the quantity plotted in Figs~(\ref{F2}) and
(\ref{F3}) is $\hat Y_{3/2}$ (for $\Delta\eta=H_{\rm I}^{-1}$). The
number of e-folds of reheating, and therefore $Y_{3/2}$, depend on the
detailed mechanism of reheating, which is unfortunately not well known
at present.  In the most standard model of reheating~\cite{KT91}, in
which the inflaton slowly decays through its coherent oscillations,
and the Universe is matter dominated, one obtains: $3N_{\rm R}\simeq
58.5 + \ln\left[(H_{\rm I}/10^{13}\,{\rm GeV})^2 (T_{\rm R}/10^9\,{\rm
GeV})^{-4}(g_{\ast}/200)^{-1}\right]$. The dilution is therefore quite
strong, and $Y_{3/2}\simeq 3\times10^{-14}\hat Y_{3/2} (H_{\rm
I}/10^{13}\,{\rm GeV}) (T_{\rm R}/10^9\, {\rm GeV})$. More generally,
if reheating takes place in an era dominated by an equation of state
of the form $p=w\rho$, the above number of e-folds is reduced by a
factor $1/(1+w)$; for $w=1/3$, for instance, which corresponds to a
relativistic fluid, one finds: $Y_{3/2}\simeq9\times10^{-8}\hat
Y_{3/2}(H_{\rm I}/10^{13}\,{\rm GeV})^{3/2} (g_{\ast s}/200)^{-1/4}$,
and the number density produced becomes independent of the reheating
temperature.  It was pointed out in Ref.~\cite{F87} that in a general
case, oscillations of an inflaton in a potential
$\sim\lambda\phi^{2n}$ would yield an equation state with $w\simeq
(n-1)/(n+1)$ after averaging out over an oscillation period. Thus, in
particular, for chaotic inflation with a potential $\lambda\phi^4$,
the Universe is indeed dominated by a relativistic fluid during
reheating ($w=1/3$).

 The ratio $Y^{\rm R}_{3/2}$ of the number density of gravitinos
produced in particle interactions during reheating to the entropy
density is, up to logarithmic corrections~\cite{M95}: $Y^{\rm
R}_{3/2}\approx 3.7\times10^{-13} (T_{\rm R}/10^9\,{\rm GeV})(g_{\ast
s}/200)^{-3/2}$. Therefore, the ratio $Y_{3/2}/Y^{\rm R}_{3/2}$ of
these two yields, assuming that reheating takes place in a matter
dominated era is: 

\begin{equation}
\frac{Y_{3/2}}{Y^{\rm R}_{3/2}}\simeq 0.1 \,\hat Y_{3/2}\,
\left(\frac{H_{\rm I}}{10^{13}\,{\rm GeV}}\right)
\left(\frac{g_{\ast s}}{200}\right)^{3/2},
\label{eq_comp}
\end{equation}

and according to Fig.~(\ref{F3}), $\hat Y_{3/2}\approx 10^{-3}
\mu^2_{1,2} (\Delta\eta/|\eta_{\rm I}|)^{-1}$, if $\mu_1=0$ and
$\mu_2=1$ or the reverse. This constitutes our main result. 
If throughout reheating, the Universe is dominated by a relativistic
equation of state, it becomes: 

\begin{equation}
\frac{Y_{3/2}}{Y^{\rm R}_{3/2}}\simeq
2\times10^5 \hat Y_{3/2} 
\left(\frac{H_{\rm I}}{10^{13}\,{\rm GeV}}\right)^{3/2} 
\left(\frac{T_{\rm R}}{10^9\,{\rm GeV}}\right)^{-1} 
\left(\frac{g_{\ast s}}{200}\right)^{5/4}.
\label{eq_comp2}
\end{equation}

Whereas the ratio of the two production yields is independent of the reheating 
temperature when the Universe is matter dominated during reheating,
it becomes inversely proportional to $T_{\rm R}$ when $w=1/3$ (relativistic
fluid). Therefore, if gravitational production is efficient, {\it i.e.},
if $m\sim H$ during or after inflation, a low reheating temperature does
not exclude a strong gravitino problem.

Let us now discuss the magnitude of $\hat Y_{3/2}$. As seen in
Fig.~(\ref{F3}), one probably has $\hat Y_{3/2}\lesssim 10^{-3}$ if
$\Delta\eta=H_{\rm I}^{-1}$, where the upper limit corresponds to
$\mu_1=0$ and $\mu_2=1$, or $\mu_1=1$ and $\mu_2=0$.  Therefore, for
reheating in a matter dominated era, one finds that the production of
gravitinos out of the vacuum is less efficient than that during
reheating, provided $\Delta\eta\gtrsim10^{-4}H_{\rm I}^{-1}$. In the
other limit, where the Universe is dominated by a relativistic equation of
state during reheating, one finds that gravitational production
can be much more efficient that reheating production of gravitinos, by a
factor $\sim 10^2 \mu^2_{1,2} (\Delta\eta/|\eta_{\rm I}|)^{-1} 
(T_{\rm R}/10^9\,{\rm GeV})^{-1}$.

Finally, an order of magnitude estimate for $\Delta\eta$ is
$\phi/\phi'$ ($\phi$ is the inflaton field) taken at the point at
which the slow-roll approximation breaks down, {\it i.e.}  where
$\dot{\phi}^2/2 \sim V(\phi)$ (a dot denotes differentiation with
respect to cosmic time). This gives $\Delta\eta \approx 2(\phi/ M_{\rm
Pl})|\eta_{\rm I}|$, with $M_{\rm Pl}\equiv (8\pi)^{1/2}m_{\rm Pl}$
the Planck mass. Therefore, $\Delta\eta\sim |\eta_{\rm I}|$ for
scenarios of the chaotic type, and $\Delta\eta\ll |\eta_{\rm I}|$ for
scenarios of the new inflation type with small field values; quite
possibly, in this latter case, $\Delta\eta<10^{-4}|\eta_{\rm
I}|$~\cite{A87}.

In new inflation, therefore, the gravitational production cannot be
neglected if $\mu_1\sim0$ and $\mu_2\sim1$ (or the reverse),
$\phi\lesssim 10^{-4}M_{\rm Pl}$ at the end of slow-roll, and $H_{\rm
I}\sim10^{13}\,$GeV.  Moreover, if reheating proceeds faster than in
the ``standard'' model (matter domination), such as in $\lambda\phi^4$
chaotic inflation, gravitational production of gravitinos can become
more efficient than reheating production. If gravitational production
dominates, cosmological bounds on the gravitino abundance at the time
of Big Bang nucleosynthesis should be turned into upper limits on the
effective mass terms of the gravitino during and after inflation, as
gravitational production is suppressed as $(m/H_{\rm I})^2$ if
$m<H_{\rm I}$.

\section{Discussion}

We discussed the conformal behavior of the gravitino in a spatially
flat FRW background spacetime. We obtained the linearized field
equation for the helicity $\pm3/2$ components, which reduces to a
Dirac like equation in curved spacetime. A massive gravitino is not
conformally invariant, and cosmological particle production ensues,
through the amplification of the vacuum fluctuations by the non-static
background metric. We assumed that the gravitino effective mass is
proportional to the Hubble constant, and used the technique of
Bogoliubov transforms to calculate the ratio $Y_{3/2}$ of the number
density of gravitinos to the entropy density at the time of
reheating. This quantity depends on the effective mass of the
gravitino during and after inflation, on the Hubble constant at the
exit of inflation ($H_{\rm I}$), on the duration of the transition
between inflation and radiation/matter domination ($\Delta\eta$), and
on the number of e-folds of reheating.  Notably, $Y_{3/2}$ scales as
the inverse of the transition timescale, which defines the degree of
``non-adiabaticity'' of the transition during which gravitinos are
produced.  The comparison of the gravitational production of
gravitinos to production in reheating depends on the details of the
mechanism of reheating, during which the gravitationally produced
gravitinos are strongly diluted. 

If we assume that the gravitino mass is of order of the Hubble
constant during or after inflation, that $H_{\rm I}\sim10^{13}\,$GeV,
and that the Universe is matter dominated throughout reheating,
gravitational production is generically less efficient than production
in reheating interactions, provided the transition is not too abrupt,
{\it i.e.}  $\Delta\eta\gtrsim10^{-4}H_{\rm I}^{-1}$. However, in
scenarios of new inflation, one can find
$\Delta\eta\lesssim10^{-4}H_{\rm I}^{-1}$, in which case gravitational
production would turn out to produce as many gravitinos, or more, than
reheating interactions. Similarly, if reheating proceeds faster, for
instance if the Universe is dominated by a relativistic fluid during
reheating, as happens in {\it e.g.}, chaotic inflation with a
potential $\lambda\phi^4$, the number density of gravitinos produced
out of the vacuum exceeds, possibly by a large factor, the density of
gravitinos produced in particle interactions during reheating. It must
be stressed that in the above, we assumed the effective mass $m$ of
the gravitino to be of order the Hubble constant, either during or
after inflation. If not, gravitational production is suppressed as
$(m/H_{\rm I})^2$.

  To conclude, let us briefly address the particular case of
pre-Big-Bang string cosmology~\cite{GV93}, in which particle
production out of the vacuum has been studied
extensively~\cite{BH98,BMUV98}, but not for spin$-3/2$. In this
scenario, one expects that, to leading order, the gravitino mass term
vanishes during inflation, if the only dynamical field is the
axion-dilaton field, as indeed, the tree level superpotential of
string-inspired supergravity does not receive contributions from the
dilaton. Similarly, $m=0$ also in the post-inflationary phase (if only
the axion-dilaton field is considered), at least until a non
perturbative superpotential for the dilaton sets in, or until
supersymmetry breaking takes place.  If we assume that the gravitino
is also massless during the so-called stringy phase, it then couples
to the axion-dilaton field only through the K\"ahler connection. The
field equation Eq.~(\ref{eq_mode3_2}) then decouples into two first
order differential equations for the mode functions $h_{ss}$ and
$g_{ss}$, whose solutions are written as in flat-space up to a
time-dependent phase which depends on the K\"ahler connection. If the
initial state corresponds to the conformal vacuum as $\eta\to-\infty$,
then $h^{\rm in}_{++}=g^{\rm in}_{--}=0$. Since we assume that the
gravitino remains massless after inflation, conformal triviality also
holds as $\eta\to+\infty$, and $h^{\rm out}_{++}=g^{\rm out}_{--}=0$.
From Eq.~(\ref{eq_Bbeta}), it is then obvious that $\beta_{ks}=0$,
{\it i.e.} no particle production takes place.

  At the next level of approximation, one should consider moduli
fields, take into account higher order corrections to the effective
action in the stringy phase, and/or introduce a non-perturbative
superpotential to stabilize the dilaton in the FRW phase. This would
lead, quite presumably, to the appearance of an effective mass for the
gravitino.  Unfortunately, these effects are difficult to implement,
because the underlying dynamics or the physics remain poorly known. To
give an example of what could be obtained, let us assume that the
gravitino is massless during the pre-Big Bang and stringy phases, and
that it acquires a mass after the exit in the FRW era. Then the
methods and results of the previous section can easily be transposed
to this scenario, since the gravitino, being massless in the pre-FRW
eras, is insensitive to the background dynamics. It is easy to verify
that if the exit in the FRW phase takes place at a scale $H_{\rm
I}\sim10^{17}\,$GeV, as has been advocated recently~\cite{BMUV98},
gravitational and reheating production of gravitinos become of the
same order, even if the Universe is matter dominated throughout
reheating, provided $m\sim H_{\rm I}$.

However, reheating in pre-Big Bang cosmology is not expected to
proceed through coherent oscillations of the ``inflaton'', and the
above estimate could turn out to be naive. A detailed study of the
mechanism of reheating in pre-Big Bang cosmology thus appears
mandatory. Depending on how fast reheating proceeds, and what
temperature is achieved, this could lead to a strong gravitino problem
(which one would naively expect if $H_{\rm I}\sim10^{17}\,$GeV), which
would thus require: $m_{3/2}\gtrsim10^4\,$GeV for an unstable
gravitino, or $m_{3/2}\lesssim 2\,$keV for a stable gravitino. A more
detailed study of this problem is left for further work.

\bigskip

\centerline{\bf Note added}\bigskip

Upon completion of this paper, we became aware of a related work by
A.~L.~Maroto and A.~Mazumdar (``Production of spin 3/2 particles from
vacuum fluctuations'', hep-ph/9904206). These authors obtained the
field equation for helicity 3/2 gravitinos, assuming
$\gamma^\mu\Psi_\mu=0$ (which projects out helicity 1/2 modes), and
calculated the amplification of vacuum fluctuations, using the
technique of Bogoliubov transforms. They applied their technique to
the production of gravitinos in preheating. In this respect, their
work and ours are complementary: gravitational production during
inflation generically produces particles with wavenumber $k\lesssim
H_{\rm I}$, while in preheating, the production takes place for modes
with $k\gtrsim H_{\rm I}$.

After the present paper was submitted, two other related studies
appeared: R.~Kallosh, L.~Kofman, A.~Linde and A.~Van~Proeyen
(``Gravitino production after inflation'', hep-ph/9907124) studied the
problem of gravitino production during inflation and during
preheating, for both helicity 1/2 and helicity 3/2 modes.  Their
important work shows that the helicity 1/2 modes are not conformally
invariant even if they are massless, and that their production in
preheating can be very large. The paper by G.~F.~Giudice,
A.~Riotto and I.~Tkachev (``Non-thermal production of dangerous relics
in the early Universe'', hep-ph/9907510) reaches similar conclusions.

\acknowledgments{It is a pleasure to thank A.~Buonanno and J.~Martin
for many valuable comments and discussions, and P.~Bin\'etruy,
R.~Brustein, B.~Carter, R.~Kallosh, A.~Linde, J.~Madore, K.~Olive,
A.~Riotto and G.~Veneziano for discussions.}

\appendix

\section*{Notations}

 We write $\gamma_\mu$ a general
relativistic Dirac matrix, and $\gamma_a=e_a^\mu\gamma_\mu$, or, if
confusion could arise, $\hat\gamma_a$, a (constant) flat-space Dirac
matrix. We define:  $\sigma_{ab}=\frac{1}{2}[\gamma_a,\gamma_b]$. The
Dirac matrices are written in the Weyl representation:

\begin{equation} 
\gamma^a = -i\left( 
\begin{array}{cc} 0 & \bar\sigma_a \\ 
\bar\sigma_a & 0 \\ 
\end{array} \right), 
\label{eq_gamma}
\end{equation} 
with: $\sigma_a=(1,\bbox{\sigma})$, and
$\bar\sigma_a=(1,-\bbox{\sigma})$, and the $\bbox{\sigma}$ are flat-space
Pauli matrices. We also define:
$\gamma_5=i\hat\gamma_0\hat\gamma_1\hat\gamma_2\hat\gamma_3$.

Our choice of vierbein for the FRW background is: $e_\mu^{{}a}=a(\eta)$,
$e^\mu_{{}a}=a(\eta)^{-1}$. The spin connection, without $\Psi$ torsion, is
then:  

\begin{equation} 
\frac{1}{4}\omega_0^{ab}\sigma_{ab}  =  0,\,\,\,\,
\frac{1}{4}\omega_i^{ab}\sigma_{ab}  =  \frac{1}{2}{\cal H}\gamma_i\gamma^0,
\label{eq_spinconnection}
\end{equation}
where we defined ${\cal H}=a'/a$.  We define the helicity operator
$\bbox{\epsilon}^{\rm L}\bbox{\sigma}$ for a Weyl spinor of momentum
$\bbox{k}$, where $\bbox{\epsilon}^{\rm L}$ is the unitary vector along
$\bbox{k}$. We then define $\chi_+(\bbox{k})$ and
$\chi_-(\bbox{k})$ as the eigenspinors of $\bbox{\epsilon}^{\rm
L}\bbox{\sigma}$ with respective helicity $+1/2$ and $-1/2$:
$\bbox{\epsilon}^{\rm L}\bbox{\sigma}\,
\chi_\pm(\bbox{k})=\pm\chi_\pm(\bbox{k})$.  We decompose a four
component spinor $\psi(\eta,\bbox{k})$ in eigenstates of
helicity~\cite{M95}, $\psi(\eta,\bbox{k}) = \psi_+(\eta,\bbox{k}) +
\psi_-(\eta,\bbox{k})$, with:

\begin{equation}
\psi_s(\eta,\bbox{k}) = 
\left( 
\begin{array}{c}
h_s(\eta)\chi_s(\bbox{k}) \\ 
sg_s(\eta)\chi_s(\bbox{k}) \\ 
\end{array}
\right),\,\,\,s=\pm, 
\end{equation} 
where $h_s(\eta)$, and $g_s(\eta)$ are (scalar) functions of conformal
time.  The eigenspinors $\chi_+$ and $\chi_-$ verify, in particular:
$\chi_s^\dagger\chi_{s'}=\delta_{ss'}$. In Section~II, we perform a
similar decomposition for the spinor-vector in terms of the mode
functions $h_{ms}(\eta)$, where $m=L,+,-$ denotes the helicity of the
polarization vector, and $s=\pm$ denotes the spinor helicity.

Finally, we define the charge conjugation operator:
$C=i\hat\gamma^2\hat\gamma^0$, and the conjugate $\psi^{\rm C}$ of a
spinor $\psi$: $\psi^{\rm C}=iC\hat\gamma^0\Psi^\ast$. One can show that:
$i\sigma_2\chi_s^\ast=-s\chi_s$, where $s=\pm$. It is then easy to show
that the conjugate of a spinor $\psi_s(\eta,-\bbox{k})$, with helicity
$s$, and momentum $-\bbox{k}$, is:

 \begin{equation}
\Psi^{\rm C}_s(\eta,-\bbox{k})= i
\left(
\begin{array}{c}
g_s^\ast\chi_{s}\\
-sh_s^\ast\chi_{s}\\
\end{array}
\right).
\label{eq_conj}
\end{equation}

This identity is useful in deriving the normalization identities of the
gravitino operator in Section~II.

\end{document}